\documentclass{egpubl}
\usepackage{eurovis2015}

\usepackage{url}
\usepackage{multirow}
\usepackage{tikz}
\usetikzlibrary{arrows,shapes,positioning,shadows,trees}
\usepackage{color}
\usepackage{colortbl}
\makeatletter 
\let\@tabclassz =\@classz 
\let\@tabclassiv =\@classiv 
\makeatother 

\usepackage{graphicx}

\title[Visualizing Cartograms: Goals and Task Taxonomy]
{
Visualizing Cartograms: Goals and Task Taxonomy\thanks{This paper is an extended version
 of~\cite{Task_C}, which appears as a short paper in the 17th Eurographics Conference on Visualization (EuroVis), 2015.}
}

\author[Nusrat and Kobourov]
{
	Sabrina Nusrat, Stephen Kobourov
	\\
	Department of Computer Science\\University of Arizona
}

\begin{document}

\teaser
{
\centering
\includegraphics[width=.27\textwidth]{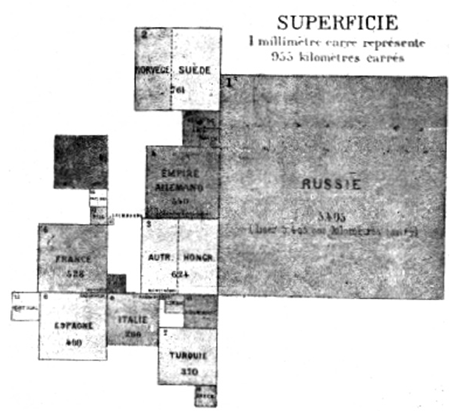}
\hspace{0.04\textwidth}
\includegraphics[width=.3\textwidth]{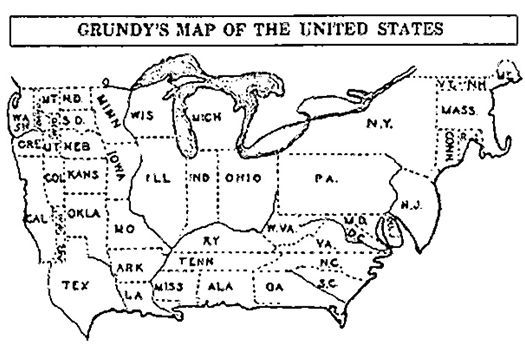}
\hspace{0.04\textwidth}
\includegraphics[width=.28\textwidth]{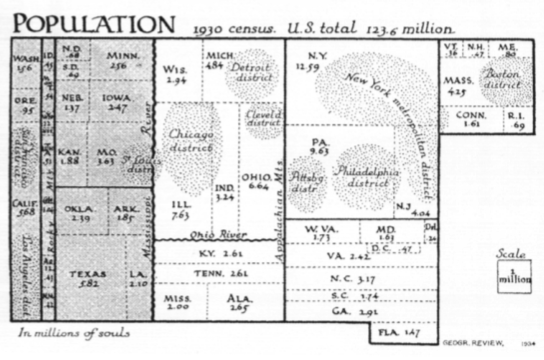}\\
\hspace{0.03\textwidth}(a)\hspace{0.35\textwidth}(b)\hspace{0.34\textwidth}(c)
 \caption{\small Early cartograms: (a) Lecasseur's 1868 "statistique figurative" map of Europe~\cite{Tobler04}; (b) Grundy's 1929 Washington Post "population and taxes" cartogram of the US~\cite{Tobler04}; (c) Raisz' rectangular population cartogram of the US~\cite{Raisz34}.}
\label{fig:cartograms}
}

\maketitle

\begin{abstract}
\thispagestyle{empty}
Cartograms are maps in which areas of geographic regions (countries, states) appear in proportion to some variable of interest (population, income).
Cartograms are popular visualizations for geo-referenced data that have been around for over a century. Newspapers, magazines, textbooks, blogs, and presentations frequently employ cartograms to show voting results, popularity, and in general, geographic patterns. Despite the popularity of cartograms and the large number of cartogram variants, there are very few studies evaluating the effectiveness of cartograms in conveying information.
In order to design cartograms as a useful visualization tool and to be able to compare the effectiveness of cartograms generated by different methods, we need to study the nature of information conveyed and the specific tasks that can be performed on cartograms. 
In this paper we consider a set of cartogram visualization tasks, based on standard taxonomies from cartography and information visualization.
We then propose a cartogram task taxonomy that can be used to organize not only the tasks considered here but also other tasks that might be added later.



\end{abstract}



\section{Introduction}

A \textit{cartogram}, or a value-by-area map, is a representation of a map where geographic regions are modified to reflect a statistic such as population or income. Geographic regions, such as countries, states and provinces of a map, are scaled by area  to visualize some statistical information, while attempting to keep the overall result readable and recognizable~\cite{KNP04, KNPS03, ks07, dorling96}. This kind of visualization has been used for many years, in fact, the first reference to the term `cartogram' dates back to at least 1868, and \'Emile Levasseur's rectangular cartograms used in an economic geography textbook~\cite{Tobler04}. Since then cartograms have been studied by geographers, cartographers, economists, social scientists, geometers, and information visualization researchers.

\begin{figure*}[htbp]
\begin{center}
\includegraphics[width=0.38\textwidth]{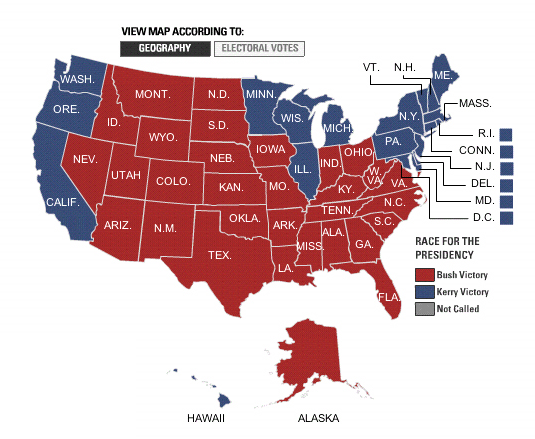}\hspace{.8cm}
\includegraphics[width=0.38\textwidth]{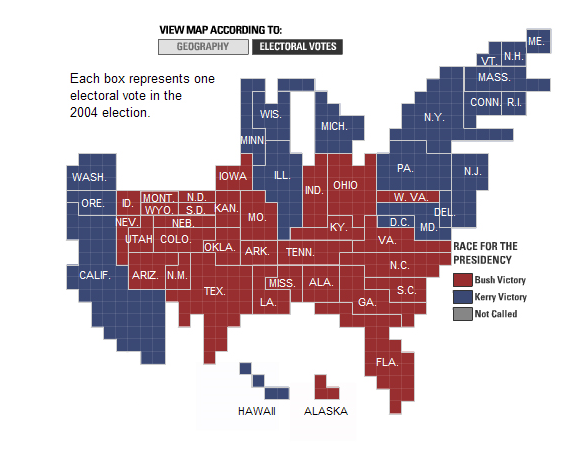}\\
(a)\hspace{0.4\textwidth}(b)
\caption{Red-blue (Republican-Democrat) map of the USA showing 2004 election results from the New York Times~\cite{NYT04}: (a) geographically accurate map,
 (b) a population cartogram.}
\label{fig:red-blue}
\end{center}
\end{figure*}

\paragraph*{Motivation:}
Given the ever-growing flood of information, cartograms provide a compact and visually appealing way to present the world's political, social and economic realities. Red-and-blue population cartograms of the United States have become an accepted standard for representing presidential election results. For example, in the 2004 election, geographically  accurate maps seemed to show an overwhelming victory for George W. Bush; while the population cartograms effectively communicated the near even split; see Fig.~\ref{fig:red-blue}. 
Likely due to aesthetic appeal and the possibility to visualize data and put political and socioeconomic reality into perspective, cartograms are widely used  in newspapers, magazines, textbooks, blogs, and presentations. For example, New York Times~\cite{NYT06} shows the election results of 2006 using some nice interactive maps and cartograms. Los Angeles Times~\cite{LAT12} follows the trend by showing 2012 election results using cartograms. 
In addition to visualizing election outcomes, cartograms are frequently used to represent other kinds of geo-referenced data. Dorling cartograms are used in the UK Guardian newspaper~\cite{Guar} to visualize social structure and in the New York Times to show the distribution of medals in the 2008 summer Olympic games~\cite{NYT_O}. Popular TED talks use cartograms to show how the news media make us perceive the world~\cite{Alisa}, to expose the myths about developing world~\cite{Hans2},
 and to visualize the complex risk factors of deadly diseases~\cite{Hans}. Cartograms continue to be used in textbooks, for example, to teach middle-school and high-school students about global demographics and human development~\cite{Class1, Class2}.

Despite the popularity of cartograms and the large number of cartogram variants, there are very few studies evaluating  cartograms. In order to design effective cartograms we need to compare cartograms generated by different methods on a variety of suitable tasks. Before such comparisons can be made, we need to understand the visualization goals and to explore the possible tasks suitable for cartograms. 
Although there is a rich literature on generating cartograms, there is very little work on evaluating the usability of cartograms and their effectiveness. In this paper we consider a set of cartogram visualization tasks, based on standard taxonomies from cartography and information visualization. We then propose a cartogram task taxonomy that can be used to organize not only the tasks considered here but also other tasks that might be added later.




\section{Related Work}

Here we survey task taxonomy work in information visualization and cartography, and we also we summarize some of many cartogram generation algorithms. 

\subsection{Task Taxonomies in Information Visualization}
Visualization tasks have been defined and classified, often depending on the context and scope of the tasks.  Wehrend~\cite{Weh} defines `visualization goals' as actions a user may perform on her data and presents nine such goals: (1) \textit{identify} (establish the characteristics by which a user can recognize an object), (2) \textit{locate} (determine the position of an object in absolute or relative terms), (3) \textit{distinguish} (recognize one object as being different than other objects), (4) \textit{categorize} (divide the set of objects into appropriate classes), (5) \textit{cluster} (group similar objects),  (6) \textit{rank} (determine the order of objects), (7) \textit{compare} (note similarities and dissimilarities in a set of objects), (8) \textit{associate} (link two or more objects based on their characteristics),  (9) \textit{correlate} (establish a relationship between two or more objects). 

Wehrend's work is extended by Zhou and Feiner~\cite{ZHOU}. They define `visualization techniques' as low-level operations and visual tasks as interfaces between high-level presentation intents and low-level visual techniques without specifying exactly `how' an operation is done. For example, if the visual presentation is intended to convey a presenter's message to a user, visual tasks that accomplish this intention are \textit{summarize} and \textit{elaborate}. Visual techniques that are used to \textit{elaborate} are \textit{emphasize} and \textit{reveal}, and examples of techniques that are used to \textit{summarize} are \textit{associate}, \textit{identify}, \textit{comapare}, and \textit{cluster}.

Whether as low-level operations or as visualization goals, visual tasks or meta-operations, \textit{identify} and \textit{compare} are listed in numerous taxonomies in cartography, HCI, GIScience and visual analytics, and their definitions are largely consistent across taxonomies (e.g., Blok et al.~\cite{BLOK}).
In addition to \textit{identify} and \textit{compare}, Maceachren~\cite{MAC} adds the task \textit{interpret} for geographic information visualization; it determines a connection between an identified feature in an abstract data representation and a real-world entity. Andreinko et al.~\cite{ANDR03} list \textit{identify} and \textit{compare} as cognitive operations for visualizing spatio-temporal data.

Some recent taxonomies do not include \textit{identify} and \textit{compare}, but rather use terminology more common in statistics. For example, Amar et al.~\cite{AMAR} present a list of low-levels tasks that capture people's activities while using information visualization tools for understanding data: (1) {\em retrieve value} (find attributes about some data case), (2) {\em filter} (find data cases satisfying some concrete attributes), (3) {\em compute derived value} (calculate an aggregate representation for a set of cases), (4) {\em find extremum} (find data cases with an extreme value of an attribute), (5) {\em sort} (rank cases according to a numeric attribute), (6) {\em determine range} (find the span of attributes for a set of cases), (7) {\em characterize distribution} (characterize the distribution of an attribute's values over the set of data cases), (8) {\em find anomalies} (identify outliers), (9) {\em cluster} (group cases based on similar attributes), and (10) {\em correlate} (identify relationships across cases by their attributes). Yi et al.~\cite{YI07} propose seven general categories of tasks widely used in interactive information visualization : 1) {\em select}, 2) {\em explore}, 3) {\em reconfigure}, 4) {\em encode}, 5) {\em abstract/elaborate}, 6) {\em filter}, and 7) {\em connect}. These represent `user intents' while interacting with a system rather than the low-level interaction techniques provided by a system. 

While the above discussion covers a general set of tasks for information visualization system designers, if the tasks are not categorized across different dimensions it becomes difficult for other systems to classify and adapt the tasks. 
The typology of abstract visualization tasks proposed by Brehmer and Munzner~\cite{BM13} focuses on three questions: {\em why} is a task performed, {\em what} are the inputs and outputs, and {\em how} is the task performed. What is particularly useful in this typology is that it distinguishes between high-level tasks (that answer why) and low-level tasks (that answer how) and provides a link between the two questions. 
Peuquet's \cite{peuquet1994s} Triad Representational Framework  store and present geographic information based on three dimensions: {\em where} (location-based),  {\em what} (object-based), and  {\em when} (time-based).
B\"orner~\cite{KATY}  considers additional important questions about what a visualization technique is developed for and how it is applied: {\em when} (in temporal analysis), {\em where} (for geospatial studies), {\em what} (topical studies) and {\em with whom} (network studies). Schulz et al.~\cite{HTMH13} address the following questions to define their design space of visualization tasks: {\em why} is a visualization task performed? {\em how} is a task carried out? {\em what} does a task seek? {\em where} in the data does a task operate? {\em when} is a task performed? {\em who} is executing a task? These questions relate to the goals of the tasks, the means, the characteristics, the target and cardinality of data entities,  the order of the tasks, and the type (expert/non-expert) audience. 


\subsection{Task Taxonomies in Cartography}

Cartography is the science and practice of making and using maps. 
%
Roth~\cite{RR13} classifies  existing taxonomical frameworks into three types:
objective-based taxonomies, operator-based taxonomies, and operand-based taxonomies.  The first type focuses on user intent, or what the user wishes to perform. Examples include \textit{identify}, \textit{compare}, and \textit{associate}. Taxonomies discussed in the previous section are mostly of this type. Operator-based taxonomies focus on operators in cartographic interfaces that supports the objective of the user. Example operators in cartographic interaction include \textit{brushing}~\cite{Shepherd, Dykes, MAC, Masters}, \textit{focusing}~\cite{BUJA, Dix, MAC, Masters, Edsall}, \textit{zooming}~\cite{Shneider96, Edsall}, and \textit{linking}~\cite{BUJA, Dix, Edsall}. In operand-based taxonomies, the focus is on the operand, or the object with which the user is interacting. Wehrend~\cite{Weh93} proposes taxonomy on seven types of data: \textit{scalar}, \textit{nominal}, \textit{direction}, \textit{shape}, \textit{position}, \textit{spatially extended region or object}, and \textit{structure}. The taxonomy of Ward and Yang~\cite{Ward} on 'interaction operands and spaces', and Persson et al.~\cite{Per06} on 'interaction types' 
are examples of operand-based taxonomies.
In the context of interactive cartography, the taxonomy provided by Andrienko et al.~\cite{ANDR03} is noteworthy for both operator-based and operand-based taxonomies. Here, the authors classify and evaluate the existing techniques and tools from the perspective of the characteristics of the spatio-temporal data they are applicable to, and the types of exploratory tasks the tools can potentially support. 
\subsection{Cartogram Generation Algorithms}

According to Tobler~\cite{Tobler04} the term `cartogram' dates back to at least 1868 and it was used to mean statistical maps, or choropleth maps~\cite{Palsky, Funk}. 
Raisz in 1934 gave a formal definition of value-by-area cartogram in~\cite{Raisz34},
although only rectangular cartograms were considered. Cartograms are studied in the information visualization literature~\cite{hkps04, FS04, KNP04, HK98} and in several  cartography textbooks~\cite{Dent, Slocum}.

\begin{figure}[htbp]
\begin{center}
\includegraphics[width=0.45\textwidth]{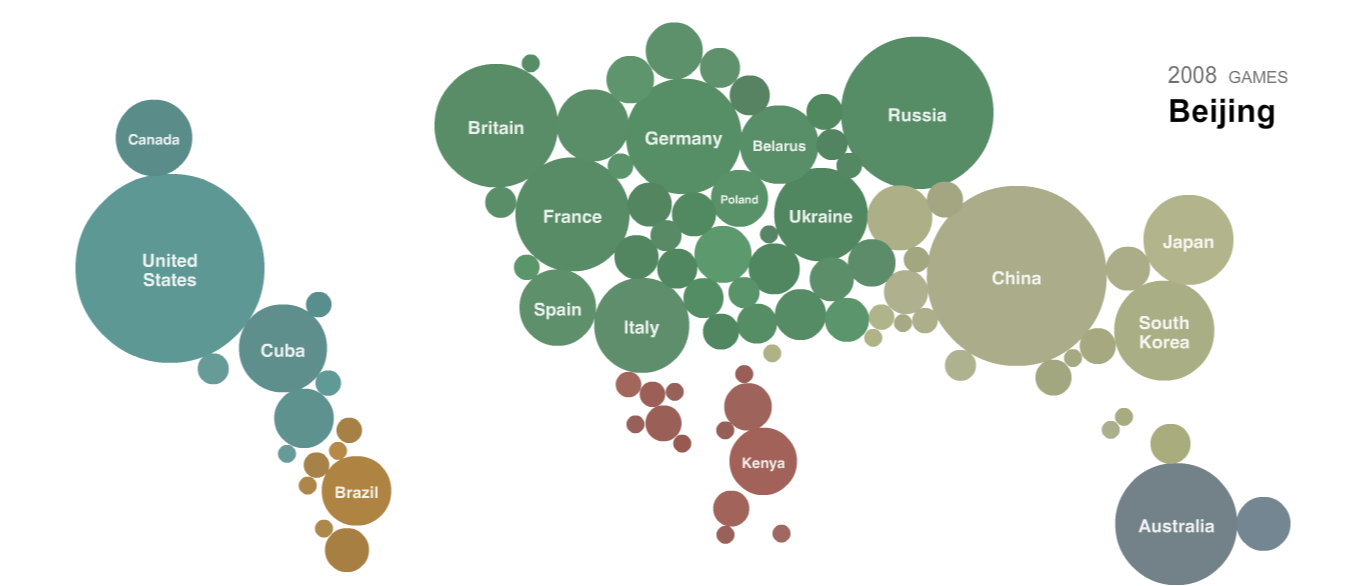}\\
(a)\\
\includegraphics[width=0.43\textwidth]{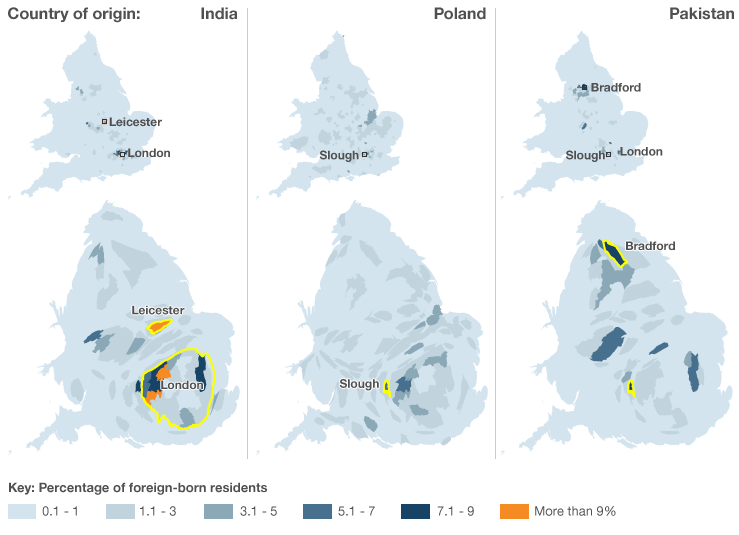}\\
(b)

\caption{(a) New York Times cartogram (Dorling) of the 2008 Olympic medals~\cite{NYT_O}; 
(b) 2012 BBC cartogram (diffusion) of migration patterns in the UK~\cite{BBC_migration}.
}
\label{fig:4Cart}
\end{center}
\end{figure} 

There is a wide variety of methods to generate cartograms, broadly categorized in four types: contiguous, non-contiguous, Dorling, and rectangular~\cite{ks07}.
In contiguous cartograms the original geographic map is modified (by pulling, pushing, and stretching the boundaries) to change the areas. Among these cartograms, the most popular method is the diffusion-based method proposed by Gastner and Newman~\cite{GN04}. 
Others of this type include the rubber-map method by Tobler~\cite{Tobler73}, contiguous cartograms by Dougenik et al.~\cite{DCN85}, 
CartoDraw by Keim et al. ~\cite{KNPS03}, constraint-based continuous cartograms by House and Kocmoud~\cite{HK98}, and medial-axis-based cartograms by Keim et al. ~\cite{KPN05}. More recent are circular arc cartograms~\cite{KKN13}.
Non-contiguous cartograms are generated by starting with the regions of the given map and scaling down each region independently, so that the desired size/area is obtained~\cite{Olson}. 
Dorling cartograms represent regions in the map by circles~\cite{Dorling91}. Data values are realized by circle size: the bigger the circle, the larger the data value. Rectangular cartograms, as their name indicates, use rectangles to represent the regions in a map. Rectangular cartograms have been used for more than 80 years~\cite{Raisz34}. More recent rectangular cartogram methods include~\cite{BSV12, ks07}.  Other topological variants include rectangular hierarchical cartograms~\cite{dykes_hierarchical} and rectilinear cartograms~\cite{BMS10,ourSoCG}. 

Two studied parameters for evaluating cartograms are {\em cartographic error} (how well do the modified areas represent the corresponding values) and {\em shape error} (how much do the modified areas resemble the originals). There have been several attempts to 
 measure the performance of existing cartogram algorithms by defining these parameters. For example, Keim et al.~\cite{KNPS03} use both cartographic error and shape error to analyze the relative performance of two algorithms: CartoDraw and VisualPoints. Buchin et al.~\cite{BSV12} also use cartographic error in the performance evaluation of rectangular cartograms.  Berg et al.~\cite{BMS10} present an algorithm for constructing rectilinear cartograms with zero cartographic error and correct region adjacencies. They compare their cartograms by the polygonal complexity (number of corners) and some measure of `fatness' of polygonal regions used. Henriques et al.~\cite{Carto_SOM} propose an algorithm Carto-SOM and compare it with some existing cartogarm generation algorithms by computing cartographic error and by visual analysis. More recently, Alam et al.~\cite{MJ_EuroVis} propose a set of quantitative measures (such as statistical distortion, topology distortion, orientation and shape distortion, and complexity) and analyze  several different types of cartograms using these measures. There is less known about cartograms in the qualitative realm. For example,  Sun et al.~\cite{Hui} measured effectiveness of different types of cartograms by ranking user preferences. Ware~\cite{Jen} performed a similar study to find the effect of animation on cartograms. 

In summary, there are numerous task taxonomies in information visualization and in cartography, but none designed specifically for cartograms. As a popular visualization method, cartograms are extensively used in the electronic and print media, but there are no comprehensive evaluations of the effectiveness of cartograms in general, and of different types of cartograms in particular.
In this paper we consider different visualization goals and tasks and propose a task taxonomy for cartograms which can be used to study the effectiveness of different types of cartograms.

\section{Task Taxonomy for Cartograms}
Although there is a large number of task taxonomies in cartography, information visualization and human-computer interaction, visualization goals and tasks are not clearly defined for cartograms. 
In this paper we address this issue by adapting existing tasks from cartography and information visualization and by adding new tasks, particularly suitable for cartograms. We categorize these tasks in four dimensions, based on the questions why, how, what, and where. We believe our list of visualization tasks and their classification can be used in formal evaluations of various cartogram generation methods. Moreover, the analysis of the goals and tasks suitable for cartograms, can have the potential to improve future cartogram design.

\subsection{Analytic Tasks and Visualization Goals}
Most cartograms are modified geographic maps which combine two features typically not present in other maps and charts: (1) they contain geographical statistical information (2) they contain location information. Therefore, cartograms have the advantage of allowing traditional cartographic tasks, as well as information visualization tasks about the encoded statistic. Through discussions with information visualization experts and using the affinity diagramming approach we put together a set of eleven tasks appropriate for cartograms. Some of these tasks are adapted from existing literature on cognitive operations, exploratory tasks, and analytic tasks in information visualization and cartography; others are particularly relevant to cartograms. Our task taxonomy does not include low-level, system-specific tasks, such as  \textit{zoom}, \textit{pan} and \textit{brush}, since we are focusing on analytic goals and tasks.
We list each of the ten tasks below, along with a general description and specific examples.

\begin{enumerate}[1.]
\item \textbf{Detect change:} This is a new task proposed for cartograms that is not present in other taxonomies. In cartograms the size of a country is changed in order to realize the input weights. Since change in size (i.e., whether a region has grown or shrunk) is a central feature of cartograms, the viewer should be able to detect such change.

\textit{Example Cartogram Task: Given a population cartogram of the USA, can the viewer detect if the state of California has grown or shrunk?}

\item \textbf{Locate:}
The task in this context corresponds to searching and finding the position of a state in a cartogram. In some taxonomies this task is denoted as \textit{locate} and in others as  \textit{lookup}. However, Brehmer and Munzner~\cite{BM13} differentiate between \textit{locate} and  \textit{lookup} tasks. In the context of cartograms, if the viewer is familiar with the USA, she can simply lookup California. On the other, hand if the viewer is unfamiliar with the USA, she has to search and locate California first. Since cartograms often drastically deform an existing map, even if the viewer is familiar with the underlying maps, finding something in the cartogram might not be a simple lookup. 

\textit{Example Cartogram Task: Given a population cartogram of the USA, locate the state of California.}

\item \textbf{Recognize:}
One of the goals in generating cartograms is to keep the original map recognizable, while distorting it to realize the given statistic. Therefore, this is an important task in our taxonomy. The aim of this task is to find out if the viewer can recognize countries/states from the original map when looking at the cartogram. 

\textit{Example Cartogram Task: Given the shape of a state from the original map and shapes of two states from the cartogram, find out which of the two cartogram states corresponds to the state from the original map.}

\item \textbf{Identify:}
The \textit{identify} task has been used in many taxonomies but conveys slightly different meanings~\cite{Weh93, BLOK, MAC}. Some authors used this task to mean geographic search in space, e.g., `identify your house based on an aerial image in Google Earth'. Identify was also used to mean temporal search, e.g., `when will the bluff erosion reach my house?' It has also been used for an attribute search, e.g., `what is the range of the endangered species?' In our taxonomy we use  \textit{identify} for attribute or characteristic search as used by  Brehmer and Munzner~\cite{BM13}. Identify focuses on a single object.

\textit{Example Cartogram Task:  If US election results are shown in a red-blue cartogram, identify the winning candidate for the state of California.}

\item \textbf{Compare:}
The  \textit{compare} task is another very commonly used one in objective-based taxonomies~\cite{RR13, MAC, Weh93}.  This task has also been used in a qualitative study of cartograms~\cite{Jen}. However,  \textit{compare} is pretty unambiguous, as the task typically asks for similarities or differences between attributes. We use is in the same way in our taxonomy.

\textit{Example Cartogram Task:  Given a population cartogram of the USA, compare two states by size.}

\item \textbf{Find top-$k$:}
This is another commonly used task in visualization. Here the goal is to find $k$ entries with the maximum
 (or minimum) values of a given attribute. This is a generic task that covers specific tasks,
 such as ``Find extremum'', where the goal is to find the data with the extreme value~\cite{AMAR}; and  ``Sort'', where all the data entries are ordered based on the value of a given attribute.

\textit{Example Cartogram Task: Given a population cartogram, find out which state has the highest/lowest population. }

\item \textbf{Filter:}
The  \textit{filter} task asks to find data cases satisfying some criteria about a given attribute, e.g, ~\cite{AMAR}. That is, the viewer can filter out examples that fail the criteria. We use this task in the same way in our taxonomy.

\textit{Example Cartogram Task: Find states which have higher population than the state of California.}

\item \textbf{Find adjacency:}
Some cartograms preserve topology, some do not. In order to understand the map characteristics properly, it is important to identify the neighboring states of a given state. Therefore, the \textit{Find adjacency} is an important new task for visualizing cartograms.

\textit{Example Cartogram Task:  Given a cartogram, find all adjacent states of California.}

\item \textbf{Cluster:}
The goal of the  \textit{cluster} task is to find objects with similar attributes. We use it in the same way for cartograms.

\textit{Example Cartogram Task:  Given a cartogram with obesity rates encoded by color, find states with have similar obesity rate as California; see Fig.~\ref{fig:4Cart}(b). }

\item \textbf{Summarize:}
Cartograms are most often used to convey a `big picture'. The \textit{summarize} task is one that asks the viewer to see the big picture. This task is associated with overviews of data and global distribution of data on the map.

\textit{Example Cartogram Task:  Given a red-blue presidential election results cartogram, determine whether it was a close election, or a "landslide win".}

\end{enumerate}
\subsection{Classification of Tasks}

We categorize the possible tasks for visualizing and interpreting information in cartograms along four dimensions: {\em goals, means, characteristics, and cardinality}; see Table~\ref{tab:dimension} for a summary. Our classification is based on three foundational typologies by Bertin~\cite{BERTIN83}, Brehmer and Munzner~\cite{BM13} and Schulz et al.~\cite{HTMH13}. In particular we use four design dimensions. 
In our taxonomy we do not differentiate between the users, as we assume that most viewers of cartograms encounter them in the popular media and hence are non-experts. Therefore, the question `who' is performing the tasks is not relevant. We also do not consider the `when' question, since most cartograms only have geo-spatial information, without any temporal component. 
With this in mind, we categorize our tasks based on the following questions:
\begin{enumerate}
\item WHY is a task performed? This identifies the objective or goal of a visualization task.
\item HOW is a task carried out? This indicates the means by which a task is carried out.
\item WHAT are data characteristics, or features of a task?
\item WHERE in the data a task operates? This relates the task with the number of instances or components a task operates on. This identifies the levels of reading for a task, whether the task or the question refers to a single element, multiple elements or all elements. 
\end{enumerate}

\end{comment}

\definecolor{Gray}{gray}{0.9}

\definecolor{LightCyan}{rgb}{0.92,0.92,1}

\definecolor{Blue}{rgb}{0.75,0.75,0.95}

\definecolor{LightB}{rgb}{0.7,0.85,1}

\definecolor{LightGreen}{rgb}{0.9,1,0.92} 

\definecolor{LightY}{rgb}{1,1,0.88}

\definecolor{LightPurple}{rgb}{1,0.95,0.95}

\definecolor{LightBlue}{rgb}{0.95,0.95,1}

\definecolor{LightPink}{rgb}{1,0.92,0.97}

\newcommand{\rot}{90}

\newcommand{\tick}
{\small 
\textcolor{black}{$\surd$}
}

\newcommand{\cross}
{\small 
\textcolor{red}{$\times$}
}

\begin{table*}[htbp]
\centering
\begin{tabular}{|c|c|c|c||c|c|c|c||c|c||c|c|c|}

\cline{2-13}

\multicolumn{1}{c|}{} &
\multicolumn{3}{c||}{\small Goals} &
\multicolumn{4}{c||}{\small Means} &
\multicolumn{2}{c||}{\small Characteristics} &
\multicolumn{3}{c|}{\small Cardinality} \\

\cline{2-13}

\multicolumn{1}{c|}{} &
\rotatebox{\rot}{\parbox{4em}{\small\centering Query}}&
\rotatebox{\rot}{\parbox{4em}{\small\centering Search}}&
\rotatebox{\rot}{\parbox{5em}{\small\centering Extract}}&
\rotatebox{\rot}{\parbox{5em}{\small\centering Map\\Relation}}&
\rotatebox{\rot}{\parbox{5em}{\small\centering Data\\Relation}}&
\rotatebox{\rot}{\parbox{5em}{\small\centering Navigation}}&
\rotatebox{\rot}{\parbox{5em}{\small\centering Derive}}&
\rotatebox{\rot}{\parbox{5em}{\small\centering Low\\Level}} &
\rotatebox{\rot}{\parbox{5em}{\small\centering High-Level}} &
\rotatebox{\rot}{\parbox{5em}{\small\centering Single}}&
\rotatebox{\rot}{\parbox{5em}{\small\centering Multiple}}&
\rotatebox{\rot}{\parbox{5em}{\small\centering All}}\\

\hline


{\small Recognize} & 
\tick &
\cross &
\cross &
\tick & 
\cross &
\cross &
\cross &
\tick &
\cross &
\tick &
\cross &
\cross \\

 {\small Detect Change}& 
\tick &
\cross &
\cross &
\tick &
\cross &
\cross &
\cross &
\tick &
\cross &
\tick &
\cross &
\cross \\

\cline{1-13}

\cline{1-13}


\rowcolor{LightCyan}
{\small Compare}& 
\tick &
\cross &
\cross &
\cross &
\tick &
\cross &
\cross &
\tick &
\cross &
\cross &
\tick &
\cross \\

\rowcolor{LightCyan}

{\small Find top-$k$}& 
\cross &
\tick &
\cross &
\cross &
\tick &
\cross &
\cross &
\cross &
\tick &
\cross &
\cross &
\tick \\

\rowcolor{LightCyan}

{\small Filter}& 
\cross &
\tick &
\cross &
\cross &
\tick &
\cross &
\cross &
\cross &
\tick &
\cross &
\cross &
\tick \\

\rowcolor{LightCyan}

{\small Cluster}& 
\cross &
\tick &
\cross &
\cross &
\tick &
\cross &
\cross &
\cross &
\tick &
\cross &
\cross &
\tick \\

\cline{1-13}

\cline{1-13}

\rowcolor{LightB}

{\small Locate}& 
\cross &
\tick &
\cross &
\cross &
\cross &
\tick &
\cross &
\tick &
\cross &
\cross &
\cross &
\tick \\

\rowcolor{LightB}

{\small Find Adjacency}& 
\cross &
\tick &
\cross &
\cross &
\cross &
\tick &
\cross &
\cross &
\tick &
\cross &
\tick &
\cross \\

\cline{1-13}

\cline{1-13}

\rowcolor{Blue}

{\small Summarize}& 
\cross &
\cross &
\tick &
\cross &
\cross &
\cross &
\tick &
\cross &
\tick &
\cross &
\cross &
\tick \\

\rowcolor{Blue}

{\small Identify}& 
\cross &
\cross &
\tick &
\cross &
\cross &
\cross &
\tick &
\tick &
\cross &
\tick &
\cross &
\cross \\
\hline



\end{tabular}\\
\caption{Tasks and their dimensions}
\label{tab:dimension}
\end{table*}


\begin{comment}
\begin{figure*}[htbp]
\centering
\scalebox{0.65}
{
\input{task_big_tree.tex}
}
\caption{Task Categorization in three dimensions}
\label{fig:tree_tasks}
\end{figure*}
\end{comment}

\begin{comment}
\begin{figure}[h]
\centering
\scalebox{0.65}
{
\input{task_category.tex}
}
\caption{Task Categorization by Goals}
\label{fig:goals}
\end{figure}
\end{comment}

\subsection{Goals}

The goal, or objective of a visualization task does not define the task itself, but rather the reason why the task being performed. We have identified five goals for cartograms.
\begin{enumerate}
\item Query: The tasks that are included here are usually local tasks; they focus on one or two objects. Some of the tasks may require comparing a state in the cartogram with the state in the original map. These tasks do not require searching through the map. Example tasks are: \textit{recognize}, \textit{detect change}. 
\item Explore: Tasks in this group require searching through the cartograms, comparing data, and finding relation among datasets. Example tasks are:  \textit{find top-$k$} and  \textit{cluster}.
\item Extract: Some tasks require extracting metadata; such tasks fall in this group. An example task is \textit{identify}.
\end{enumerate}


\begin{comment}
\begin{figure}[h]
\centering
\scalebox{0.65}
{
\input{task_category2.tex}
}
\caption{Task Categorization by Means}
\label{fig:means}
\end{figure}
\end{comment}

\subsection{Means}
The means of visualization tasks do not define the tasks themselves, but rather explain how the tasks can be performed~\cite{HTMH13}.  We have identified three different means.

\begin{enumerate}
\item Navigation: One of the methods for performing visualization tasks is to navigate or browse through the dataset. Example navigation tasks are: \textit{locate, find adjacency}.
\item Relation: This includes all means to find some relation (e.g., similarity or difference).
For cartograms, we further subdivide Relation into:
\begin{enumerate}
\item Relation across geography or data-relation: these require finding a relation in the data. Example data relation tasks are: \textit{compare, cluster}.
\item Relation across visualization or map-relation: these require finding a relation between the original map and the cartogram. Example data relation tasks are: \textit{recognize, detect change}.

\end{enumerate} 
\item Derive: The tasks in this group are performed by extraction of information, or abstraction of the data. This often involves augmentation, reduction, or filtering of data. Example derive tasks are  \textit{identify} and  \textit{summarize}.

\end{enumerate} 

\subsection{Characteristics}

This dimension does not define the task itself, rather identifies what is the level of complexity of the visualization task. Characteristics or features of a visualization task depend highly on the type of information that the task aims to reveal~\cite{HTMH13}. In the context of cartograms, these characteristics can be divided into two categories:
\begin{enumerate}
\item Low-level data characteristics: involve simple tasks that can be performed by observation from the visualization. Example tasks for cartograms are: \textit{identify, locate, compare}.
\item High-level data characteristics: involve more complex tasks that need to be deduced from the visualization. Example tasks for cartograms include: \textit{filter, cluster, sort, find top-$k$}.
\end{enumerate}


\subsection{Cardinality}
The cardinality of a task specifies where the task operates. This dimension directly relates tasks with the components of data. The reading levels by Bertin~\cite{BERTIN83} contain three types: \textit{elementary}, \textit{intermediate} and \textit{overall}, and they deal with a single data element, multiple  elements and all elements, respectively. Similar differentiation is made by Schulz et al.~\cite{HTMH13} and Yi et al.~\cite{YI10}.
Thus, the cardinality of a cartogram task differentiates  the number of regions that are investigated by a task: a single region, multiple regions, or the entire map. 
Example tasks that consider a \textit{single instance} are: \textit{identify, detect change, recognize}.
Example tasks that consider multiple instances are: \textit{compare, find adjacency}. Example tasks that require \textit{all instances} are \textit{summarize}, \textit{cluster}. 


\section {Discussion and Limitations}
Cartograms represent geo-referenced data, promote visual thinking, and transform data into stories. 
Our proposed task taxonomy focuses on the tasks that are suitable for cartograms, and is informed by earlier taxonomies for information visualization and cartography.

\begin{figure}[t]
\centering
\includegraphics[width=0.5\textwidth]{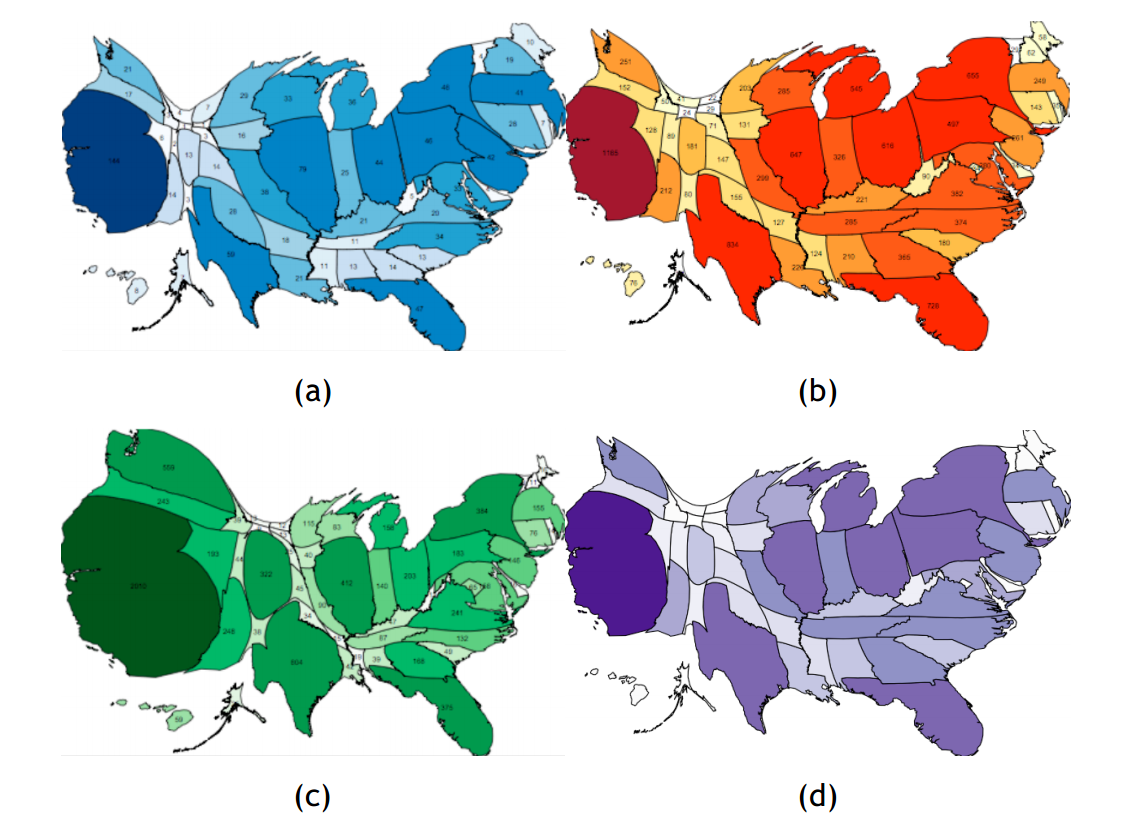}
\caption{(a) Walmart cartogram. (b) McDonalds cartogram. (c) Starbucks cartogram. (d) State population cartogram~\cite{Starbucks}.}
\label{fig:starbucks}
\end{figure}

Even though cartograms have been used for more than a century,
 there are criticisms on the informative value of cartograms~\cite{dorling96, Andy_blog}.
 There could be fundamental flaws (such as encoding size with color progression or with difference in areas) or poor design that leads to misinterpretation of the data. For example, Steven's power law~\cite{Steven_law} indicates that length has better correlation with magnitude than area or color. 
The distortion of shapes in many cartograms (e.g., rectangular) make it hard to recognize the geography of locations. 
However, unlike bar graphs (which represent size better), cartograms contain geographic information and adjacency relations. This makes it possible to see broader patterns and trends. 
In addition, cartograms are capable of mapping more than one variable (for e.g., population with size, winning candidate with color in Fig.~\ref{fig:red-blue}). These are non-trivial advantages that make it possible to provide better overview and `big-picture' summary of the underlying data.

\begin{figure}[t]
\centering
\includegraphics[width=0.32\textwidth]{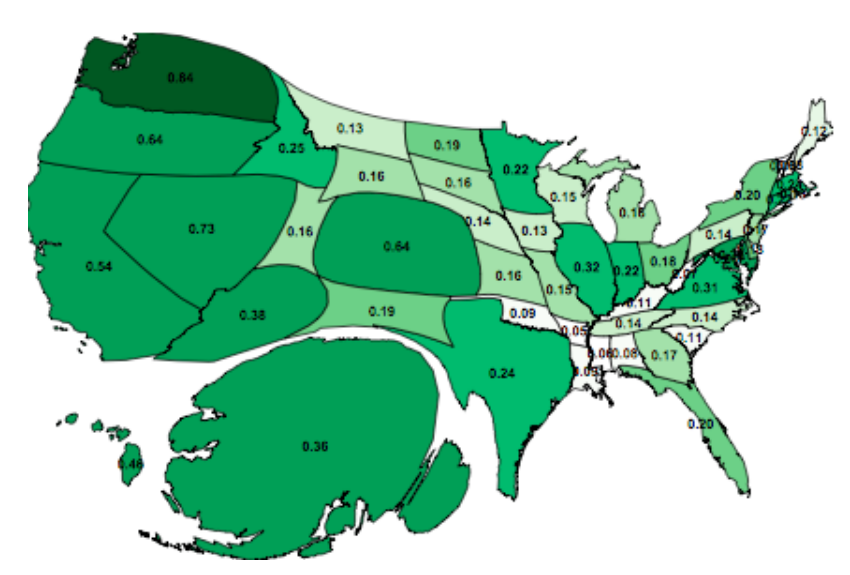}
\caption{Per capita Starbucks cartogram~\cite{Starbucks}.}
\label{fig:pc_starbucks}
\end{figure}

Cartograms are capable of representing complex data, for example, two or more cartograms can be used to correlate data and to find hidden patterns. Consider the cartograms showing Walmarts, McDonalds restaurants, Starbucks cafes, and population~\cite{Starbucks} shown in Figure~\ref{fig:starbucks}. The Walmart and McDonalds cartograms are very similar to the population cartogram, indicating that Walmart and Mcdonalds are where people are. On the other hand, the Starbucks cartogram shows that the coffee shops have a higher density in the west coast: Washington, California, Arizona, Nevada. In fact, the per capita cartogram in Fig.~\ref{fig:pc_starbucks} confirms  that Washington state (the home of Starbucks) has highest rate. Comparing the raw-numbers cartogram in Figure~\ref{fig:starbucks}(c) with the per-capita cartogram in Figure~\ref{fig:pc_starbucks} highlights the danger of misinterpretation. For example, California and Utah have very similar values per capita, but very different values in the raw-numbers cartogram. 

Our proposed taxonomy does not address the issues of misinterpretation, or the possible flaws of encoding size with area or color. However, the taxonomy can be used to evaluate different types of cartograms and determine how serious these flaws and misinterpretations are. Our taxonomy could be extended by considering lower level, system-specific tasks (e.g., zooming, linking).  However, we aimed for generality and believe that the proposed taxonomy can be applied to most types of cartograms (deformation, topological, Dorling, non-contiguous).
We did not consider tasks about dynamic and interactive cartogram visualization systems.
There can also be compound tasks that consist of two or more basic tasks.

As in other taxonomies, there are tasks that are compound and depend on simpler tasks. 
For example, we have tasks that are ``low-level'' and tasks that are ``high-level''. In order to pursue high-level tasks (e.g.  ``sort'') we often need to perform multiple low-level tasks (e.g., ``compare'').


Based on existing taxonomies from cartography and information visualization, we propose a taxonomy specifically designed with cartograms in mind.
We categorize tasks in multiple dimensions that can be useful in the evaluation of different types of cartograms. For example, the first two tasks, ``recognize'' and ``detect change'', have similar goals, means, characteristics and cardinality as they both deal with the shapes and sizes of regions; see Table~\ref{tab:dimension}. 
Based on similar patterns we have grouped ``compare'', ``find top-$k$'', ``filter'', and ``cluster'' as they deal with size comparison. Finally,  ``locate'' and ``find adjacency'' form a group and ``summarize'' and ``identify'' form a group. 

Given the many different types of cartograms, it is impossible to impose uniform cartogram requirements, but a comprehensive collection of tasks should make a fair evaluation possible. 
 To cover a variety of cartogram-specific tasks in such an evaluation, it would suffice to pick one task from each of the four groups, but a thorough evaluation will require at least seven tasks (as the last three groups have two distinct goals/means/characteristics/cardinality patterns). 
 While a single taxonomy is rarely complete and covers all possible tasks and task dimensions, the proposed taxonomy can be a useful guideline for the design and evaluation of cartograms and we have recently used it in an evaluation of four major types of cartograms~\cite{InfoVis_cartogram}.

\begin{comment}
\section{Conclusion and Future work}

In this paper we identified a set of tasks that are suitable for cartograms.
We then proposed a task taxonomy for cartograms, based on four design dimensions: goals, means, characteristics, cardinality. We plan to use this taxonomy in a user-study to evaluate the performance of different types of cartograms. We would also like to formally evaluate the impact of the use of color progression and change in area to encode differences in size, as well as misinterpretations such as those that arise from encoding raw-data or per capita data. 

\end{comment}


\bibliographystyle{eg-alpha}


\newcommand{\etalchar}[1]{$^{#1}$}


\end{document}